\documentclass[rnote]{aa}

\usepackage{txfonts}
\usepackage{graphicx}
\usepackage{natbib} 
\bibpunct{(}{)}{;}{a}{}{,}

\begin{document}

\title{Universality of the companion mass-ratio distribution}

\author{M. Reggiani\inst{\ref{inst1}} \and M. R. Meyer\inst{\ref{inst1}}}

\institute{Institute of Astronomy, ETH Zurich, CH-8093 Zurich, Switzerland\label{inst1}}

\abstract 
{}
{We present new results regarding the companion mass-ratio distribution (CMRD) of stars, as a follow-up of our previous work. }
{We used a maximum-likelihood-estimation method to re-derive the field CMRD power law avoiding dependence on the arbitrary binning. We also  considered two new surveys of multiples in the field for solar-type stars and M dwarfs to test the universality of the CMRD.} 
{We found no significant differences in the CMRD for M dwarfs and solar-type stars compared with previous results over the common mass ratio and separation range. The new best-fit power law of the CMRD in the field, combining two previous sets of data, is $dN/dq \propto q^{\beta}$, with $\beta=0.25\pm0.29$.}  
{}

\keywords{binaries: general - stars: formation - stars: low-mass - stars: solar-type}
\titlerunning{Universality of the CMRD} 
\authorrunning{M. Reggiani and M. R. Meyer}
\maketitle
% =========================
% 1. INTRODUCTION
% =========================

\section{Introduction}
A large portion of stars, both in the field \citep{Raghavan2010,Janson2012} and in star-forming regions \citep{Patience2002}, are formed in multiple systems. Therefore understanding multiple star formation is necessary to investigate star formation in general \citep{Goodwin2007,Duchene2013}.
Because binary properties reflect the main characteristics of binary formation, they may help us determining the most common mechanisms for the formation of multiple stars.
In a binary system of stars with masses $M_{1}$ and $M_{2}$ ($M_{1}>M_{2}$), the mass.ratio is conventionally defined as $q=M_{2}/M_{1}$. Similar to the initial mass function (IMF) for single objects, the companion mass-ratio distribution (CMRD) is the distribution of $q$ values  as a function of primary mass.
Tidal capture models predict that for each primary star the mass of the secondary is chosen randomly from the single-star mass function, and the CMRD reflects the IMF \citep{McDonald1993,Kroupa2003}. 
In fragmentation scenarios subsequent continued accretion onto both objects from a common reservoir tends to equalize the masses, resulting in a $q$ distribution peaked toward unity \citep{Bate2000}.
Capture is unlikely to be the most relevant binary formation mechanism, but it may still occur during the dissolution phase of star clusters, causing differences in the shape of the CMRD as a function of orbital separation \citep{Moeckel2010,Moeckel2011}.
Motivated by the fact that every theoretical model predicts a different shape of the mass-ratio distribution and of dependency of the CMRD on the primary mass, we used Monte-Carlo simulations to compare the CMRD for different samples and to study the relationship between the IMF and the CMRD (\cite{Reggiani2011}, hereafter, RM11).
This research note represents a follow-up to RM11, in which we reanalyze the "universal" CMRD by adopting a different statistical approach (Section \ref{sec:1}) and some new results on the CMRD on the basis of recent datasets (Section \ref{sec:2}).

% =========================
% 2. 
% =========================

\section{Universal companion mass-ratio distribution} \label{sec:1}
The CMRD appears to be universal over a wide range of $q$ values and primary masses \citep[e.g.][]{Metchev2009}.
According to RM11, the CMRD follows a single-slope power law $dN/dq \propto q^{\beta}$ over the separation range 1-2400 AU and primary mass range 0.25-6.5 M$_{\odot}$, and there is no evidence for variation of the CMRD with orbital separation. 

In previous work we combined samples of M dwarfs \citep{Fischer1992} and G stars \citep{Metchev2009} in the field and intermediate mass stars in ScoOB2 \citep{Kouwenhoven2005} adopting a $\chi^2$  fit of the combined binned distribution to derive the power law slope, obtaining $\beta=-0.50\pm0.29$ \citep{Reggiani2011}.
The choice of the statistical method was motivated by the need of comparing our results with previous studies of the CMRD \citep[e.g.][]{Kouwenhoven2005,Metchev2009}. However, the $\chi^2$  fit of a binned distribution can lead to a biased estimate, in particular for small samples.
A more robust analysis is instead achieved through a maximum-likelihood estimation method \citep{Feigelson2011}.
This approach gives a new best-fit power law $dN/dq \propto q^{\beta}$, with $\beta=-0.18\pm0.33$ to the data described in RM11.
Although the two values are consistent with each other within the errors, the new estimate is flatter than previously thought.

% =========================
% 3. 
% =========================

\section{Updates to the CMRD in the field} \label{sec:2}
Recently, two new studies of the CMRD for solar-type \citep{Raghavan2010} and M-dwarf primaries \citep{Janson2012} in the field have been carried out.
Since they represent the most complete samples to date for sun-like stars and M dwarfs, respectively, we applied the same statistical analysis as was presented in RM11 to follow up this preliminary work.

In the first study \citep{Raghavan2010}, roughly 200 binaries with primary masses between 0.5-3 M$_{\odot}$ were considered to determine the CMRD over a wide range of separations ($10^{-1}$-$10^5$ AU) and mass ratios (0.02-1).
The new CMRD appears to be more peaked toward unity than previously observed and the period distribution is unimodal and roughly log-normal with a peak at around 50 AU. 
Following the methodology described in RM11, we used a KS test to compare the newly observed CMRD with the CMRD by \cite{Metchev2009}, over the common range of mass ratios (0.02-1) and separations (28-1590 AU). The KS test returns a probability of $\sim30\%$, therefore we cannot reject the hypothesis that the data were drawn from the same parent population.
However, when we compare the two samples over the common range of mass ratios, irrespective of separation, the probability is only $\sim1\%$, pointing toward a change of the CMRD with orbital radius, because that of \cite{Raghavan2010} covers a wider range than that of \cite{Metchev2009}.
We therefore tested the possibility of a variation of the CMRD with angular separation. To do this we considered break points in the angular separation distribution between $10^{-1}$ and $10^5$ AU and used a KS test for each of them to determine the probability that the CMRD inside the break point is consistent with the CMRD outside. Because we found probabilities greater than 1\% for any possible choice of break point, we conclude that we have no strong evidence for a dependence of the CMRD on angular separation. 
Moreover, because we do not expect to see random pairing from cluster dissolution models inside $10^4$ AU \citep{Kouwenhoven2010} and these widest binaries are relatively rare, we need larger samples in the future to test these models.

The second study \citep{Janson2012} consists of 85 systems with primary masses between 0.15-0.5 M$_{\odot}$, separations in the range 3-227 AU and mass ratios between 0.1 and 1. For M dwarfs, the CMRD appears to be flat and the period distribution is narrower and peaks at lower values than for solar type primaries, indicating a continuous transition from higher- to lower-mass stars \citep{Burgasser2007}. 
In this case as well, we tested the newly observed CMRD with the CMRD from \cite{Fischer1992} over the common range of mass ratios and separations.
With a probability of $\sim$56$\%$ the KS test does not allow us to reject the hypothesis that the newer data were drawn from the same parent sample.
Finally, we used the same procedure as we adopted for sun-like stars, but in the range 3-227 AU, to explore the dependence of the observed CMRD on separation. We saw no evidence of this dependence either for this sample. 

%\begin{figure}
%\includegraphics[scale=0.38]{./fig1.eps}
%\includegraphics[scale=0.38]{./fig2.eps}
%\caption{\small{From top to bottom the comparison between Monte-Carlo simulations of the stellar CMRD and the observations is shown for solar type primaries and M dwarf primaries in the field respectively. The hatched histogram represents the observed CMRD for each dataset. Superimposed with a dashed line is the CMRD generated adopting the CMRD fit described in Section \ref{sec:1}, whereas with a dotted line is a Monte-Carlo simulation of a flat CMRD.}}\label{fig:1}
%\end{figure}

\begin{figure}
\includegraphics[scale=0.38]{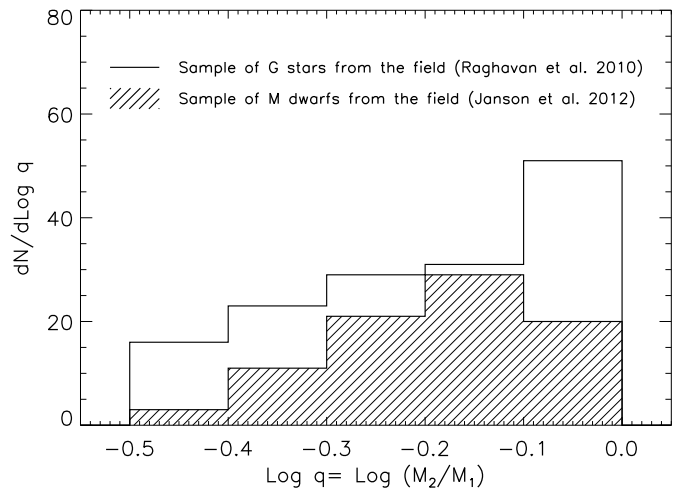}
\caption{\small{Comparison between the observed CMRDs for solar-type primaries and M-dwarf primaries in the field. The open histogram represents the CMRD from \cite{Raghavan2010}, whereas the hashed histogram represents the distribution from \cite{Janson2012}. The KS test returns a probability of $30\%$ that the two distributions are drawn from the same parent sample.}}\label{fig:1}
\end{figure}

Moreover, we compared the CMRD for solar-type primaries from \cite{Raghavan2010} with the new sample of M-dwarf primary binaries \citep{Janson2012}.
The KS test returned a probability of $30\%$ that the two distributions are consistent with each other (Figure \ref{fig:1}).
Motivated by this result and because the CMRD is independent of angular separation, we combined the two CMRDs over the common range of mass ratios. We again used a maximum-likelihood method to fit the distribution and found a power law $dN/dq \propto q^{\beta}$, with $\beta$=0.25$\pm$0.29. While this slope $\beta$ is formally consistent with the one derived in Section \ref{sec:1} (within the errors), the change in sign is significant. It is also worth mentioning that this fit is consistent with the mass-ratio distribution with power-law exponent $\beta$=-0.10$\pm$0.58 presented in a recent study of O-type spectroscopic binaries \citep{Sana2012}, whereas the observed CMRD for brown dwarfs ($\beta\sim$1.5) points toward a different formation mechanisms for these objects \citep{Goodwin2013}. 

% =========================
% 3. CONCLUSIONS
% =========================
\section{Summary} \label{conclusion}
In this research note we have presented some updates to the study of RM11.
First, we adopted a maximum-likelihood estimation method to re-derive the field CMRD power law, based on the combination of samples \citep{Fischer1992, Metchev2009, Kouwenhoven2005} described in RM11, to show how the dependence on the bin size can bias the result.

Secondly, we analyzed recent binarity studies from the field \citep{Raghavan2010,Janson2012} adopting the same methodology as in RM11.
The new results from \cite{Raghavan2010} appear to be consistent with \cite{Metchev2009} over the common range of mass ratios and angular separations. The recent updates on the M-dwarf CMRD \citep{Janson2012} are also consistent with past results.
The KS test does not allow us to reject the hypothesis that the CMRDs from \cite{Raghavan2010} and \cite{Janson2012} are drawn from the same parent sample. In both studies we uncovered no evidence for a dependence of the CMRD on separation. Therefore we combined the two distributions and obtained a new maximum-likelihood fit to the field CMRD $dN/dq \propto q^{\beta}$, with $\beta$=0.25$\pm$0.29.

Since the CMRD appears to be independent of separation and dynamical evolution \citep[see also][]{Parker2013}, it represents a measurable parameter of binary stars to focus on when investigating binary formation mechanisms.
However, we need larger samples to look for subtle variations of the CMRD with separation.
In the  future we aim to study the CMRD in other star-forming regions (e.g. the ONC) and  test its dependence on separation for wide systems.

\begin{acknowledgements} 
We thank the referee, Simon Goodwin, for his review. We are also grateful to Carolina Bergfors, Stanimir Metchev, Deepak Raghavan, Eric Feigelson, and Richard Parker for the insightful discussions and sharing their data electronically. 

\end{acknowledgements}

% =========================
% THE BIBLIOGRAPHY
% =========================

\clearpage


\begin{thebibliography}{}

\bibitem[Bate(2000)]{Bate2000} Bate, M.~R.\ 2000, \mnras, 314, 33
\bibitem[Burgasser et al.(2007)]{Burgasser2007} Burgasser, A.~J., 
Reid, I.~N., Siegler, N., et al.\ 2007, Protostars and Planets V, 427 
\bibitem[Duch{\^e}ne 
\& Kraus(2013)]{Duchene2013} Duch{\^e}ne, G., \& Kraus, A.\ 2013, arXiv:1303.3028 
\bibitem[Feigelson \& Babu(2011)]{Feigelson2011} Feigelson, E. D. \& Babu, G. J. 2011, Modern Statistical Methods for Astronomy
with R Applications, Cambridge Univ. Press
\bibitem[Fischer \& Marcy(1992)]{Fischer1992} Fischer, D.~A., \& Marcy, G.~W.\ 1992, \apj, 396, 178
\bibitem[Goodwin(2013)]{Goodwin2013} Goodwin, S.~P.\ 2013, \mnras, 
430, L6 
\bibitem[Goodwin et al.(2007)]{Goodwin2007} Goodwin, S.~P., Kroupa, 
P., Goodman, A., \& Burkert, A.\ 2007, Protostars and Planets V, 133 
\bibitem[Kouwenhoven et al.(2005)]{Kouwenhoven2005} Kouwenhoven, M.~B.~N., Brown, A.~G.~A., Zinnecker, H., Kaper, L., \& Portegies Zwart, S.~F.\ 2005, \aap, 430, 137
\bibitem[Kouwenhoven et al.(2010)]{Kouwenhoven2010} Kouwenhoven, 
M.~B.~N., Goodwin, S.~P., Parker, R.~J., et al.\ 2010, \mnras, 404, 1835
\bibitem[Kroupa et al.(2003)]{Kroupa2003} Kroupa, P., Bouvier, J., 
Duch{\^e}ne, G., \& Moraux, E.\ 2003, \mnras, 346, 354 
\bibitem[Janson et al.(2012)]{Janson2012} Janson, M., Hormuth, F., 
Bergfors, C., et al.\ 2012, \apj, 754, 44 
\bibitem[McDonald
\& Clarke(1993)]{McDonald1993} McDonald, J.~M., \& Clarke, C.~J.\ 1993, \mnras, 262, 800
\bibitem[Metchev \& Hillenbrand(2009)]{Metchev2009} Metchev, S.~A., \& Hillenbrand, L.~A.\ 2009, \apjs, 181, 62
\bibitem[Moeckel 
\& Bate(2010)]{Moeckel2010} Moeckel, N., \& Bate, M.~R.\ 2010, \mnras, 404, 721 
\bibitem[Moeckel 
\& Clarke(2011)]{Moeckel2011} Moeckel, N., \& Clarke, C.~J.\ 2011, \mnras, 415, 1179 
\bibitem[Parker 
\& Reggiani(2013)]{Parker2013} Parker, R.~J., \& Reggiani, M.~M.\ 2013, arXiv:1304.3123 
\bibitem[Patience et al.(2002)]{Patience2002} Patience, J., Ghez, 
A.~M., Reid, I.~N., \& Matthews, K.\ 2002, \aj, 123, 1570 
\bibitem[Raghavan et al.(2010)]{Raghavan2010} Raghavan, D., 
McAlister, H.~A., Henry, T.~J., et al.\ 2010, \apjs, 190, 1 
\bibitem[Reggiani 
\& Meyer(2011)]{Reggiani2011} Reggiani, M.~M., \& Meyer, M.~R.\ 2011, \apj, 738, 60
\bibitem[Sana et al.(2012)]{Sana2012} Sana, H., de Mink, S.~E., 
de Koter, A., et al.\ 2012, Science, 337, 444 


\end{thebibliography}
\end{document}